\documentstyle[12pt]{article}
\newcounter{num}
\begin{document}
\baselineskip 16pt plus 2pt
\hfill TECHNION-PH-35 \par
\hfill hep-th/9212069 \par
\hfill December 1992

\begin{center}
{\bf STABILISED MATRIX MODELS FOR NON-PERTURBATIVE} \\
{\bf TWO DIMENSIONAL QUANTUM GRAVITY} \\

{\bf Joshua Feinberg} \footnote{Supported in part by the Henri Gutwirth
Fund}\footnote{Bitnet:
PHR74JF@TECHNION.BITNET}

{\it Department of Physics, Technion-Israel Institute of Technology,} \\
{\it Haifa 32000, Israel}
\vspace{0.5cm}

\end{center}
\begin{abstract}

{\rm A thorough analysis of stochastically stabilised hermitian one matrix
models for two dimensional quantum gravity at all its $(2,2k-1)$
multicritical points is made. It is stressed that only the zero
fermion sector of the supersymmetric hamiltonian, i.e., the forward
Fokker-Planck hamiltonian, is relevant for the analysis of bosonic
matter coupled to two dimensional gravity. Therefore, supersymmetry
breaking is not the physical mechanism that creates non perturbative
effects in the case of points of even multicriticality  $k$. Non
perturbative effects in the string coupling constant $g_{str}$ result
in a loss of any explicit relation to the KdV hierarchy equations
in the latter case, while maintaining the perturbative genus
expansion. As a by-product of our analysis it is explicitly proved
that polynomials orthogonal relative to an arbitrary weight
$\exp (-\beta V(x))$ along the whole real line obey an Hartree-Fock
equation.}
\end{abstract}
\pagebreak

\noindent {\bf I.\ \ Introduction} \\

Zero dimensional hermitian one matrix models describe at their
double scaling multicritical points minimal non unitary conformal matter
coupled to two dimensional quantum gravity \cite{GrMg,BDShS,DgSh,BrKz,Br,Mg}.
Points of even multicriticality $k$, and in particular, pure gravity
$(k=2)$, exhibit non-perturbative ambiguities, instabilities and violations of
the Schwinger-Dyson (loop) equations\cite{DgSeiSh,Sh,GinZJ,FD1,FD2,FD3,FD4,Mg}.

The source of all these inconsistencies is the fact that the
critical matrix potentials of even multicriticality  $k$ are bottomless.
Thus, it is conceivable that a sensible definition of these
models should correspond to well-defined stabilised matrix
potentials.

Bottomless matrix potentials occur also in multimatrix models\cite{Dg}
describing unitary minimal conformal matter coupled to two
dimensional quantum gravity such as the two matrix model
corresponding to the Ising case\cite{GrMg,BrDgKzSh,Br}.
Thus, the problem of stabilisation is associated not only with
one matrix models and their non-unitary minimal conformal
content.

An attempt to stabilise the matrix integral,defining the partition
function of the original hermitian matrix model by rotating
integration contours in the complex eigenvalue plane
in the case of pure gravity $(k=2)$,corresponds to a unique
complex solution of the string equation free of singularities
along the real cosmological constant axis\cite{FD2,FD3,FD4}.
This implies complex susceptibility in
the matrix model which is physically unacceptable. These
facts generalise to higher even $k$ multicriticalities\cite{ChLk}.

Marinari and Parisi\cite{MP}
have suggested a possible way out of this difficulty in the
case of pure gravity by supersymmetrising the model.
The zero fermion sector in ref.\cite{MP}
is also the forward Fokker-Planck hamiltonian associated with the
Langevin equation whose force term is
$-U^{\prime}(\Phi)$,
where
$U(\Phi)$,
is the original zero dimensional matrix action.
Therefore, the definition of pure two dimensional quantum gravity
in \cite{MP}
is equivalent to the stabilisation procedure developed in \cite{GrHal}
for bottomless actions, as far as the zero fermion supersymmetry
sector of the former is concerned.

Specialising to a cubic superpotential (corresponding to a cubic
$U(\Phi)$),
Marinari and Parisi have reduced their model in the zero fermion
sector to a one dimensional non interacting Fermi gas in an
external potential.

The stabilised pure gravity model of ref.\cite{MP}
was further analysed in refs.\cite{KM,Mg}
where the one eigenvalue double scaled hamiltonian was extracted,
and a non-perturbative ambiguity free analysis of the density
of particles and energy levels was made. Numerical analysis
of the puncture one point function in this model was carried out in \cite{AJV}
where a comparison with the formulation of
\cite{FD2,FD3,FD4}
has been given.
A rather general discussion of the stochastic formulation as a
natural framework for two dimensional quantum gravity and issues
of universality in the case of pure gravity may be found in \cite{MSGV}.
The issue of non-perturbative supersymmetry breaking in the
model of ref.\cite{MP}
has been investigated in \cite{AD}.

The generalisation of the Marinari-Parisi model to higher
multicritical points
$k \geq 3$
is hard to analyse, even in the zero fermion sector of the
supersymmetric hamiltonian, since the latter corresponds to an
interacting one dimensional Fermi gas, with long range two body
interactions. Only for odd multicriticalities
$k = 3,5,7,...$
whose matrix potentials
$U(\Phi)$
are stable, is the ground state of this gas given exactly
by the Slater determinant of the first $N$ orthonormal polynomials
corresponding to
$U(\Phi)$
(multiplied by the appropriate weight factors),
in accordance with \cite{GrHal,KM}.

A semiclassical analysis of this gas, using Hartree-Fock methods,
has been carried out in \cite{MSG}.
In this work it has been found that the effective one body potential
gives rise to a particle (i.e. eigenvalue) density which is
identical to that of the original zero-dimensional matrix model
\cite{DgSh,Neu,BIZ,BPIZ,Kaz},
thereby showing that this interacting gas has a $k$th order
multicritical behaviour.

We have independently obtained similar results in \cite{JF}
by analysing the ``bosonic'' analogue
\footnote{in the sense of using symmetric rather than antisymmetric
functions of the matrix eigenvalues} of \cite{MSG},
namely - by constructing the leading $1/N$ term in the collective
field formulation
\cite{Sak,DJ}
of the Fokker-Planck matrix hamiltonian.

An attempt to construct a collective field theory for the
zero dimensional multicritical one matrix models, has been made in
\cite{JCDA}.
In that paper {\em ad hoc} arguments were invoked to justify the  use
of a collective field jacobian
identical
to the one used in \cite{DJ}
for the $c=1$ theory. Such a jacobian is needed in order to obtain
non vanishing connected multiloop correlators in the collective
field formalism. In our construction \cite{JF}
we have obtained this jacobian automatically, because our matrix
$\Phi$
depends on the Langevin time and therefore the matrix kinetic
energy term in the Fokker-Planck hamiltonian is the same as
the one considered by \cite{DJ}.
Another ansatz for a collective field formulations of the  zero
dimensional
hermitian one matrix model, that can be used to obtain the correct
loop equations, has been given in
\cite{Jev}.

The bosonic collective field theory of \cite{Sak,DJ}
has been supersymmetrised in   \cite{JR} in
an attempt to build a field theory for one dimensional superstrings,
analogous to the construction of   \cite{DJ}
for the bosonic string field theory. This theory has been applied in
\cite{JVT,RVT}
to the supersymmetric Marinari-Parisi matrix action. Its zero fermion
sector coincides with our construction in \cite{JF}.

An additional interesting approach to eigenvalue dynamics of zero dimensional
hermitian one matrix models different from the collective field
methods described in the references above, but closely related
to them is the so-called ``Conjugate Field Theory'' of   \cite{SBM}.

As will be shown later in this paper - the KdV hierarchy structure
of the zero dimensional matrix model
\cite{GrMg,DgSh,BDShS,Loop}
is completely lost under its stochastic stabilisation, if the
original matrix potential were bottomless, i.e., at the even
$k$ multicritical points.

An alternative non-perturbatively consistent definition for
$(2,2k-1)$ conformal matter coupled to two dimensional quantum
gravity, which preserves the KdV structure mentioned above is given
in  \cite{DALLEY}.
This formulation is realised in terms of the so-called ``complex
matrix'' models, in which only positive semidefinite hermitean
matrices
are included in the matrix integral that defines
the summation over surfaces
\cite{MORRIS}.

The stochastic definition for quantum gravity and the definition
mentioned above have been compared in the case of pure gravity
$(k=2)$ in \cite{AJM}
and in the first reference of \cite{DALLEY}.
A notable difference between the two definitions has been observed
already on the sphere for negative cosmological constant, namely,
the stochastic definition yielded a string susceptibility scaling
like
$(-t)^{1/2}$,
$t$ being the renormalised cosmological constant, while the definition
of  \cite{DALLEY}
implied that that susceptibility should have vanished asymptotically
as
$t \rightarrow -\infty$.
The exact susceptibilities were also compared in the references
mentioned above. This difference between the two susceptibilities
implies at once that the stochasticly defined model violates the
KdV flows non perturbatively. We will give a general proof of
this behaviour below.\\

The paper is organised as follows:  \\
In section (II) we describe in detail the stochastic stabilisation
for matrix models and state explicit justification for its use.
In Section (III) we stress the vacuous role played by supersymmetry
and its spontaneous breakdown in these models, by considering matrix
potentials that do not break supersymmetry. A possible connection
between these potentials and the Penner model is conjectured.
Section (IV) is devoted to a brief summary of our work \cite{JF}
where we have analysed the stochastic models on the sphere and
demonstrated their correct multicritical behaviour. We also
add some new remarks on this subject.   Some details related to
this section are deferred to the appendix at the end of the paper.

In section (V) we prove that the KdV structure at even
multicriticalities is completely lost by the stochastic
stabilisation. Section (VI) is devoted to a proof of the
fact that polynomials orthonormal with respect to an arbitrary
weight along the whole real line obey a system of Hartree-Fock
equations. This forms a new result to our best knowledge, in
the theory of orthogonal polynomials. We draw our conclusions
in section (VII).
\pagebreak

{\bf II. \ Stochastic Stabilisation of Matrix Integrals} \\

The forward Fokker-Planck hamiltonian, used in refs.
\cite{MP,AJV,KM}
for the pure gravity case, reads for a general matrix potential
$U(\Phi)$
\begin{eqnarray}
{\cal H} = \frac{1}{2} {\rm Tr} \left[ \left(
-\frac{\partial}{\partial \Phi} + \frac{1}{2} \beta U^{\prime}
(\Phi) \right) \cdot
\left( \frac{\partial}{\partial \Phi} + \frac{1}{2} \beta
U^{\prime} (\Phi) \right) \right] \ .       
\end{eqnarray}

Here $\Phi$ is an hermitian
$N \times N$
matrix, depending on the Langevin time coordinate (i.e. the bosonic
coordinate of the superspace used in ref.\cite{MP}).
${\cal H}$ in Eq.\ (1) is a well-defined
Schr\"{o}dinger operator. Its potential is clearly bounded from
below and grows to plus infinity as the matrix  eigenvalues
become infinite.

Therefore, a well-defined unique
($U(N)$
singlet) normalisable ground state vector
$\Psi_0(\Phi)$
exists. This fact must cure the inconsistencies of zero dimensional
matrix models described in the introduction. Note that if
$U(\Phi)$
is bounded from below such that the Boltzmann weight of the
zero-dimensional matrix model is normalisable this ground state
is given by
\begin{eqnarray}
&& \Psi_0 (\Phi) = \frac{1}{\sqrt{Z}} \exp[ - \frac{1}{2} \beta
\ {\rm Tr} U(\Phi) \ , \nonumber \\
&& Z = \int d^{N^2} \Phi \exp [ - \beta \ {\rm Tr} U(\Phi) ] \ .  
\end{eqnarray}
This state is also the {\it zero energy} ground state of the zero
fermionic sector of the full supersymmetric hamiltonian used in
\cite{MP} and supersymmetry {\it is not} broken. Moreover, expectation
values of operators, all at infinite Langevin time project only
onto the ground  state $\Psi_0$, and are identical to the corresponding
correlators in the original zero-dimensional matrix model:
\begin{eqnarray}
\langle\Psi_0|{\cal O}_1(\Phi)\dots{\cal O}_n(\Phi)|\Psi_0\rangle =
\frac{1}{Z} \int {\rm d}^{N^2}\Phi \exp[-\beta{\rm Tr} U(\Phi)]
{\cal O}_1(\Phi)\dots {\cal O}_n(\Phi) \ .               
\end{eqnarray}

If, however,  $U(\Phi)$ is unbounded from below, the zero-dimensional
Boltzmann
weight is unnormalizable and the corresponding matrix model exists only
at a saddle point level.  The ground state in the zero fermion sector
$\Psi_0$ of
the supersymmetric hamiltonian has positive energy $E_0$.  This does not
imply supersummetry breaking, since a zero energy ground state may be
found in another supersymmetric sector of the full supersymmetric
hamiltonian generalising the one used by \cite{MP} to arbitrary matrix
potentials.  We will comment more on this in the next section.
Alternatively, the appropriate Langevin equation has in this case
runaway solutions only, and the Fokker-Planck probability density at any
finite portion of the matrix eigenvalue space decays asymptotically in
Langevin time $t$ as $\exp(-E_ot)$\cite{PaBk,ZJBk}\footnote{For this
property to hold also in the double scaling limit, we must ensure that
the vacuum remains non-degenerate even as $N \rightarrow \infty$, i.e.,
that the energy eigenvalue $E_1$ of the first excited state of {\cal H}
does not coalesce with $E_0$.  As was shown in ref.\cite{MP} the mass
gap $E_1-E_0$ double scales in the WKB approximation for $k=2$.
Moreover, we will show that it double scales for any value of $k$.  Thus
the vacuum state remains non-degenerate.  This was tacitly assumed in
Ref.\cite{GrHal}}.\footnote{An attempt to compactify runaway
solutions of the Langevin equation has been carried out in ref.\cite{MPL}
but will not be discussed here.}

However, averages of operators over the weight defined by the diagonal
(euclidean) evolution kernel of the supersymmetric hamiltonian,
normalised by the average of the unit operator are well defined as the
Langevin time goes to infinity, and correspond to
\begin{eqnarray}
\langle{\cal O}_1(\Phi)\dots {\cal O}_n(\Phi) \rangle_{t\rightarrow\infty}
= \int {\rm d}^{N^2}\Phi|\Psi_0(\Phi)|^2{\cal O}_1(\Phi)\dots{\cal
O}_n(\Phi)|_{t=\infty} \ .                        
\end{eqnarray}
Here $\Psi_0(\Phi)$ is the normalisable ground state of {\cal H}, and
all the operators on the RHS of eq.\ (4) are at $t=\infty$.  We thus
consider eq.\ (4) as the stabilised definition for correlators in the
case of bottomless matrix potentials.  Clearly eq.\ (4) yields real
results for hermitean operators ${\cal O}_i$. Therefore stabilised
correlators of scaling operators in the original matrix models are real.
In particular, the two puncture correlator, identified as the random
surface specific heat in the original matrix models, must be real and
positive.

As was argued in \cite{GrHal}, the asymptotic perturbative expansion of
correlators of the form given in eq.\ (4) as power series in the non
linear couplings of the matrix potential $U(\Phi)$ is identical to the
corresponding expansion in the original unstable matrix integrals in
their domain of validity as a saddle point evaluation of the diverging
matrix integral.  Therefore, upon reorganising these series into a $1/N$
expansion, these series should agree order by order also in the genus
expansion as long as the couplings are in the saddle point domain.  In
the next section we will demonstrate this fact for the leading planar terms.

As is well known, the laplacian over hermitean matrices acquires the
form
\begin{eqnarray}
-{\rm Tr} \frac{\partial^2}{\partial \Phi^2} =
-\frac{1}{\Delta(x)} \sum^N_{i=1} \frac{\partial^2}{\partial x^2_i}
\Delta (x)
+ \left( {\rm U}(N)~{\rm angular~momentum~terms}\right) \  
\end{eqnarray}
where $x_i$ are the matrix eigenvalues and $\Delta(x_i)$ is the
Vandermonde determinant.  This leads to the mapping of eigenvalue
dynamics onto that of a one-dimensional Fermi gas\cite{BPIZ}.  Clearly,
the ground state $\Psi_0(\Phi)$ mentioned above is a U$(N)$ singlet.

For a generic potential $U(\Phi)$, the hamiltonian in eq.\ (1) contains
 long range two body interaction terms among the eigenvalues if the
degree of $U(\Phi)$ is higher than cubic:
\begin{eqnarray}
{\cal H}_{\rm int} =
-\frac{1}{4} \beta {\rm Tr} \left[\left(\frac{\partial}{\partial
\Phi}\right) U^\prime(\Phi)\right] =
-\frac{1}{4} \beta \sum_{i,j}
\frac{U^\prime(x_i)-U^\prime(x_j)}{x_i - x_j} \ .     
\end{eqnarray}
Thus, generally, the one dimensional gas of eigenvalues is an
interacting Fermi gas.  This fact, together with the definition of
stabilised correlators in eqs.\ (3) and (4) as Fermi gas Green's
functions, synchronous in Langevin time $t$, make the stochastic
stabilisation a most natural definition of two dimensional quantum
gravity.  Indeed\cite{JF}, the authors of \cite{BDShS} have postulated
the existence of a Fermi gas hamiltonian to facilitate computations of
connected Green's functions of loop operators by using Wick's theorem.
They needed such an hamiltonian to define time ordered products of loop
operators.  The $N$ body ground state of that hamiltonian has been
recognised as the Slater determinant of the first $N$ orthogonal
polynomials $P_n(x)$ of\cite{GrMg,DgSh,BrKz} (with the appropriate
$\exp[-\frac{1}{2} \beta U(x)]$ weights) and the gravitational connected
multi-loop correlators were obtained as the synchronous limit of the
Green's functions mentioned above\footnote{It was stated incorrectly
in \cite{BDShS} that the hamiltonian mentioned above should have been a
``one body hamiltonian whose eigenstates are the
$P_n(x)\exp[-\frac{1}{2}\beta U(x)]$''.  It is well known that the only
one dimensional one body hamiltonian defined along the whole real line,
whose square integrable eigenstates form a set of polynomials orthogonal
relative to a weight is the harmonic oscillator hamiltonian.  The corresponding
polynomials are, of course, the Hermite polynomials.  It is clear that
the arguments given in \cite{BDShS} remain intact if the ground state of
the Fermi gas hamiltonian is the Slater determinant described above, as
is the case there.}.

Naturality of the stochastic definition of matrix models for two
dimensional quantum gravity appears also as one analyses the genus
expansion by using collective field techniques.  This will be discussed
in the next section.

The interaction terms in eq.\ (6) may have an interesting
interpretation\cite{JF}:\\
If TrU$^\prime(\Phi)$ is expanded as $\sum_{n\geq 0}c_n{\rm Tr}\Phi^n$
we obtain ${\cal H}_{\rm int} = \sum_{n \geq 1} c_n \sum^{n-1}_{\ell=0}
{\rm Tr}\Phi^{\ell}\\
&&~ {\rm Tr}\Phi^{n-\ell-1}$.  Therefore, from the point
of view of the (non critical) one dimensional matrix theory ${\cal H}_0 =
\frac{1}{2}[-\frac{\partial^2}{\partial\phi^2} +
\frac{1}{4}\beta^2U^\prime(\Phi)^2]$, whose eigenvalues form a
non-interacting Fermi gas in the singlet sector, ${\cal H}_{\rm int}$
may be interpretted as higher curvature terms\cite{HCT} that push the
system to its multicritical point.

In the case of pure gravity $(k = 2)$ where one uses a cubic matrix
potential $U(\Phi)$, the interaction term of eq.\ (6) reduces to an
interaction of the puncture operator ${\rm Tr}\Phi$ with a constant
background
potential $\sim {\rm Tr} {\bf\large 1} = N$ proportional to the number
of particles\cite{MP,KM,JF} and the Fermi gas may be analysed as a
noninteracting one.  This gas. however, may be considered only as a
canonical ensemble, since the number of fermions in the gas cannot be
changed without changing the shape of the effective potential\cite{JF}.
Therefore, the associated Fermi energy must be evaluated
self-consistently from the N-dependent effective potential\cite{MP}
which is also the condition that the eigenvalue distribution of the
matrix will be supported along a real single segment\cite{FD3} (at least
in the planar approximation and for positive cosmological constant).
Therefore, unlike the case of the $d=1$ model\cite{DJ,GrKl}, the Fermi
energy is not a free parameter that can be used to define the double
scaling limit.  This observation is certainly true for higher
multicritical potential since ${\cal H}_{\rm int}$ in eq.\ (6) depends
explicitly on $N$.
\pagebreak

\noindent{\bf III. The (no) Role of Supersymmetry and Its Breakdown}\\

The supersymmetric matrix model in ref.\cite{MP} has been interpretted
by its authors as a summation method over supersurfaces embedded in a
one dimensional supersymmetric target space -- i.e., as a (one
dimensional) superstring.

Supersymmetry in this model is spontaneously broken, since the
superpotential is cubic\cite{Witten}.  The strength of this effect is
non-perturbatively small in the string coupling constant for positive
cosmological constant\cite{MP,AD} but appears already on the sphere for
negative cosmological constant\cite{MP,FD3,AJM}.
Various authors have argued that nonperturbative supersymmetry breaking
effects in this ``superstring'' may serve as a source of nonperturbative
gravitational effects or as toy models for supersymmetry breaking in
realistic superstring models\cite{DgSeiSh,AD}.
A supercollective field analysis of the super matrix model of\cite{MP}
was carried out in \cite{JVT,RVT} using the formalism developed
in \cite{JR}, in order to gain a better understanding of ``target
space'' supersymmetry of that superstring.  This analysis resulted in a
conclusion that one could not obtain a superstring field theory
analogous to the string field theory obtained from \cite{DJ}.  In
particular, no extra dimension has been seen to emerge in contrast ot
the case in \cite{DJ}, since the ``time of flight'' in the semiclassical
collective analysis (i.e., the inverse mass gap of single particle
excitations\cite{JF}) was finite\cite{JVT,RVT}. Moreover, multicritical
behaviour appears only at the zero-fermion supersymmetry sector, with
$(2,2k-1$) singularities\cite{JF}.
This means that the ``superstring'' of \cite{MP} has no clear target
superspace interpretation.

We interpret these outcomes as a signature that the supersymmetric model
of \cite{MP} does not describe a discretisation of a system containing
genuine superconformal matter coupled to random supergeometry.  These
results rather stress that this ``superstring'' is equivalent to the
coupling of ($2,2k-1$) conformal matter to ordinary random worldsheet
geometry, and that the roles of supersymmetry and, in particular, its
spontaneous breakdown for even values of $k$, are vacuous.  This must
hold at least as long as $(2,2k-1)$ multicritical one matrix potentials
$U(\Phi)$ are concerned in eq.\ (1).\footnote{In order to obtain a
superstring field theory analogous to \cite{DJ}, one must first
construct
a supermatrix model for $\hat{c}=3/2$ superconformal matter coupled to
super Liouville theory.I thank J.P.Rodrigues for correspondence on this
point.}

We prove our claim that supersymmetry and its beakdown play no role in
the stochastic definition of two dimensional quantum gravity, described
in the previous section by considering even multicritical matrix
potentials $U_k(\Phi) = U_k(-\Phi)$ describing the $(2,2k-1)$
multicritical point in eq.\ (1).  For even multicriticalities $k$, the
$U_k(\Phi)$ are bottomless.  In these cases, the full supersymmetric
matrix hamiltonian has a zero energy supersymmetric ground state in its
$N^2$-fermionic sector\cite{Witten}, and supersymmetry is unbroken.
Specifically, the $N^2$-fermionic sector hamiltonian is
\begin{eqnarray}
{\cal H} = \frac{1}{2} {\rm Tr} \left[\left(\frac{\partial}{\partial
\Phi} + \frac{1}{2} \beta U^\prime(\Phi)\right) \cdot
\left(-\frac{\partial}{\partial\Phi} + \frac{1}{2}
\beta U^\prime(\Phi)\right)\right]                   
\end{eqnarray}
and the corresponding ground state is proportional to
\begin{eqnarray}
\Psi_0^{N^2}(\Phi) \sim \exp [+\frac{1}{2} \beta {\rm Tr} U(\Phi)] \ .
\end{eqnarray}

Thus, if one considers the stabilised matrix model as a
genuine supersymmetric model, supersymmetry dictates using the potential
$-U_k(\Phi)$, rather than $+U_k(\Phi)$, as a poligonation Boltzmann
weight at even multicriticalities $k$ \ \footnote{We have supressed in
eq.\
(8) the trivial $N^2$ fermion Fock-space factor, which decouples form
correlation functions of the form similar to that of eqs.\ (3) and (4),
as long as zero-fermionic number operators ${\cal O}_i(\Phi)$ are
considered} which is clearly not critical and may support multiband
eigenvalue distributions.  Thus, for the class of even potentials
considered above, supersymmetry is never broken, and $(2,2k-1)$
multicriticality and the associated non perturbative gravitational
effects are associated only with the zero fermion sector of the
supersymmetric hamiltonian\cite{JF}.  The latter may include the
exact zero energy supersymmetric ground state (for odd values of $k$),
or may not (for even values of $k$).

The fact that only the zero fermionic sector of the supersymmetric
matrix model is relevant, as far as non perturbative two dimensional
quantum gravity is concerned, implies that in this respect the
supersymmetric stabilisation of the gravitational system degenerates
into the forward Fokker-Planck formulation described in the previous
section.

We close this section by making a comment regarding the noncritical
potentials $-U_k(\Phi)$ discussed above (where $k$ is either even or
odd).  These potentials may be made critical by adding to them a
$\ln\Phi^2$ piece:
\begin{eqnarray}
\tilde{U}_k(\Phi) = - U_k(\Phi) + \ln \Phi^2          
\end{eqnarray}
for which we find on the sphere\cite{GrMg}
\begin{eqnarray}
\tilde{W}_k(R) = \oint \frac{dz}{2\pi i} \tilde{U}^\prime_k (z +
\frac{R}{z} ) = 1 +(1-R)^k               \ .        
\end{eqnarray}
Here $R(x)$   (which has the usual meaning\cite{GrMg,DgSh,BrKz}) is
given by
\begin{eqnarray}
R(x) = 1 - (x-1)^{1/k} \ .               
\end{eqnarray}
Upon double scaling (following the notations of\cite{GrMg})
\begin{eqnarray}
1 - x = t \beta^{- 2k/2k+1} \ , \ \
1-R(x) = f(t)\beta^{-2/2k+1}       
\end{eqnarray}
we find that the spherical double scaled susceptibility is
\begin{eqnarray}
f(t) = (-t)^{1/k}    \ .             
\end{eqnarray}
Therefore, $\tilde{U}_k(\Phi)$ leads to a $k$-th order multicritical
behaviour, with sign flipped cosmological constant.

It would be interesting to check\cite{JF1} whether these $(2,2k-1)$
multicritical points may be connected to the $c=1$ Penner
model\cite{DIVA,CDL,CIT,DMP} via a phase transition reducing the central
charge to non positive values\footnote{We are indebted to C.\
\v{C}rnkovi\'{c} and to R.\ Plesser for  discussions on this point.}.\\
Indeed, it seems that the $(2,2k-1)$ phase corresponds to a single
eigenvalue band located inside the logarithmic well, while in the $c=1$
Penner phase we have multiband eigenvalue support in shallow minima to
the sides of the logarithmic well.
\pagebreak

\noindent{\bf IV. \ Collective Field Analysis on the Sphere}\\

In order to study the stabilised multicritical points with
multicriticality $k \geq 3$, one has to cope with the interaction term
among eigenvalues in eq.\ (6).  Here we will discuss scaling properties
of these models and show that they are identical to those of the
original ones.  Since only the $U(N)$-singlet ground state
$\Psi_0(\Phi)$
of ${\cal H}$ is involved, it is natural to analyse the interacting gas
(in
the spherical approximation) in terms of the fermion density operator --
i.e., the collective field $\phi(x)$ associated with the matrix $\Phi$.

Following\cite{Sak,DJ} we define\footnote{We have used the normalisation
of ref.\cite{GrKl}} the collective field as
\begin{eqnarray}
\phi(x) = \frac{1}{\beta} {\rm Tr} \delta ({\bf 1} \cdot x - \Phi) =
\frac{1}{\beta} \sum^N_{i=1} \delta(x-x_i)          
\end{eqnarray}
which implies that $\phi(x)$ is a non-negative operator and obeys the
normalisation condition
\begin{eqnarray}
\int \phi(x) dx = N/\beta \ .
\end{eqnarray}

Since $\Phi$ depends on the Langevin  time $t$ and its dynamics is
fixed
by ${\cal H}$ in eq.\ (1), the jacobian of the trnaformation from the
matrix eigenvalue variables to the collective field is identical to
the one used in the $d=1$ case\cite{DJ}.  This immediate
conclusion, stemming form the very definition of the stabilised model,
bypasses the need to invoke {\em ad hoc} arguments of the type used
in \cite{JCDA} or postulates about the form of the zero dimensional
collective field partition function as in ref.\cite{Jev}, that are
needed if one makes the tranformation to collective modes directly in
the zero-dimensional matrix model.
Therefore, also from the collective field point of view, the stochastic
definition of two dimensional quantum gravity seems natural, as we have
argued before.

In this section we concentrate on the spherical approximation to ${\cal
H}$ in order to demonstrate that the $k$-th order multicritical
behaviour is respected by our formalism.  The planar collective field
action $S_0[\phi]=\beta^3\int{\cal L}_{(0)}dxdt$ for the matrix
hamiltonian ${\cal H}$ in eq.\ (1) is given by\cite{Sak,DJ}:
\begin{eqnarray}
{\cal L}_{(0)}&&= - \{ \frac{\pi^2}{6} \phi^3(x) +
[\frac{1}{8} U^{\prime^2}(x) - \mu_F] \phi(x) \}+ \nonumber \\
&&\frac{1}{4} \int dy \frac{U^\prime(x)-U^\prime(y)}{x-y}
\phi(x)\phi(y) \                                              
\end{eqnarray}
as we have discussed in \cite{JF}.

Here $\mu_F$ is a lagrange multiplier (the chemical potential) that
enforces the constraint of eq.\ (15).  Unlike the case of $d=1$
matter\cite{DJ}, $\mu_F$ here is not a free parameter whose deviation
from
a critical value (which implies a change in the number of fermions) is
used to define the scaling behaviour, in accordance with the discussion
at the end of section (II).

The planar collective field equation of motion is readily found to be
\begin{eqnarray}
&&-\frac{\delta S[\phi]}{\delta\phi(x)}|_{\rm planar} =
\frac{1}{2}\pi^2\phi(x)^2 + \frac{1}{8}U^\prime(x)^2-\mu_F \nonumber \\
&&~~~ - \frac{1}{2} \int {\rm d}y \frac{U^\prime(x) - U^\prime(y)}{x-y}
\phi(y) = 0 \ ,            
\end{eqnarray}
where $\phi(x)$ is subjected to the constraint of eq.\ (15), and that by
definition, $\phi(x)$ is non-negative.

A crucial observation is that for $N/\beta = 1$ and $\mu_F = 0$, this
non-linear non-trivial integral equation is identical to the planar
limit of the Schwinger-Dyson equation obeyed by the loop operator in
the original zero-dimensional matrix model\cite{Neu,BPIZ,BIZ}:
\begin{eqnarray}
&&F(z)^2 - U^\prime(x)F(z)+\eta(z) = 0 \ ,      \nonumber \\
&&F(z) = \lim_{\beta \approx N \rightarrow\infty}
\langle \frac{1}{\beta} {\rm Tr} \frac{1}{z-\Phi} \rangle \ , \nonumber
\\
&&\eta(z) = \lim_{\beta\approx N\rightarrow\infty}
\langle\frac{1}{\beta} {\rm Tr}
\frac{U^\prime(z) - U^\prime(\Phi)}{z-\Phi} \rangle \ ,  
\end{eqnarray}
when $z$ approaches the real axis\footnote{As $z=x-{\rm i}\epsilon, \
\epsilon \rightarrow 0+$ we have $F(z) = \frac{1}{2}U^\prime(x) + {\rm
i}\pi\phi(x)$ thus making the imaginary part of eq.\ (18) vanishing and
its real part proportional to eq.\ (17).}.  The fact that eq.\ (17) is
identical to the planar loop equation of the original matrix model is
not surprising and conforms with the postulates of ref.\cite{GrHal}.
Moreover, it seems that the WKB expansion of eqs.\ (3) and (4) should
correspond term by term to the genus expansion of the corresponding
Schwinger-Dyson (i.e. -  loop) equations in the original model as we
have argued in section (II).

Therefore, under the conditions $N/\beta=1$ and $\mu_F = 0, \ \phi(x)$
that solves eq.\ (17) is just the planar limit eigenvalue density of the
original Dyson gas in an external potential $U(x)$.

Thus, for a matrix potential $U(\Phi)$ in the universality class of the
$k$th multicritical point, $\phi(x)$ will exhibit $k$th order
multicritical behaviour.

Solutions to eq.\ (18) are usually taken such that $\phi(x)$ is
supported along a given real segment where $F(z)$ has a cut (consider
e.g. \cite{Neu}) and scaling perturbations that are introduced do not
change this structure.  However, since we are not free to change $N$
(i.e., to shift $\mu_F$ in eq.\ (17)) in our formalism we cannot
introduce a cosmological constant and its conjugate puncture operator to
our formalism in this way.  Therefore, in order to allow for a
cosmological constant perturbation, we let one end point of the support
of $\phi_k(x)$ which solves eq.\ (17) with $\mu_F = 0$ precisely at the
$k$th multicritical point to vary on a proper scale\footnote{In the
$d=1$ matrix model\cite{DJ}, variation of the chemical potential $\mu_F$
changes the location of the classical turning points which are the end
points of supp$\{\phi(x)\}$.}\cite{JF}.
This scale must be that of the double scaled fluctuations of the matrix
near its critical point (which we take as $\Phi_c=2)$, i.e.,
$\beta^{-2/(2k+1)}$.
Therefore, the required eigenvalue distribution should be supported
along a segment $[-2,b]$, where
\begin{eqnarray}
b = 2 - \epsilon \beta^{-2/(2k+1)}             
\end{eqnarray}
in which $\epsilon$ is the double scaled specific heat.

Such a deformation of $\phi_k(x)$ alone is not enough to obtain the
desired scaling behaviour of $1-N/\beta$, since it generically induces
all $k-1$ relevant deformations\cite{GrMg,Neu} present at the $k$th
multicritical point.  The desired solution to eqs.\ (15) and (17) must
therefore include counterterms that will cancel these unwanted scaling
contributions to $1-N/\beta$.

In \cite{JF} we have found it to be given by
\begin{eqnarray}
\phi(x) = &&\frac{1}{\pi}\left[(2+x)(b-x)\right]^{\frac{1}{2}}
\{ C_k(b-x)^{k-1} + \nonumber \\
&& \sum^{k-1}_{n=1} \beta^{-2(k-n)/(2k+1)}
\left( \matrix{k + 1 \cr n + 1} \right) \left(\frac{\epsilon}{4}
\right)^{k-n} \cdot C_n(b-x)^{n-1} \}        
\end{eqnarray}
where $C_n = n!(n+1)!/2(2n)!$ are normalisation constants.  The
corresponding deformed matrix potential is given by \cite{Neu}
\begin{eqnarray}
U^\prime(x) &=&
2\pi\{[(2+x)(x-b)]^{\frac{1}{2}} \cdot
[C_k(x-b)^{k-1} + \nonumber \\
&& \sum^{k-1}_{n=1} \beta^{-2(k-n)/(2k+1)}
\left( \matrix{k + 1 \cr n + 1} \right) \left(\frac{\epsilon}{4}
\right)^{k-n} C_n(x-b)^{n-1}]\}_+    \ . 
\end{eqnarray}
{}From eq.\ (15) we find
\begin{eqnarray}
\frac{N}{\beta} = 1 - (k+1) \left(\frac{\epsilon}{4}\right)^k
\beta^{-2k/(2k+1)} \cdot
\left(1+{\cal O}(\beta^{-2/(2k+1)} \right)  \ .     \nonumber
\end{eqnarray}
Therefore, the renormalised cosmological constant, defined as
$\beta^{2k/(2k+1)}(1-\frac{N}{\beta})$, is given by\footnote{The
numerical
coefficients in front of powers of $\epsilon$ in eqs. (20)-(22) were
found in closed from by M.\ Moshe.
} $t = (k+1)(\frac{\epsilon}{4})^k$, or
equivalently, the susceptibility is
\begin{eqnarray}
\epsilon = 4 (\frac{t}{k+1})^{1/k} \ .     
\end{eqnarray}
Therefore, the stabilised model retain its $k$-th order multicritical
scaling behaviour on the sphere.

As in the original zero matrix models, the susceptibility becomes
complex for $t<0$ (i.e. in the $N>\beta$ pahse) at even
multicriticalities $k$.  This, in turn, implies that $\phi(x)$ in eq.\
(20) becomes supported along a single segment in the complex eigenvalue
plane, where it is complex.  Such a behaviour is unacceptable since the
coordinates of the fermions in the gas are real by definition.  Thus the
appearance of a complex solution for $\phi(x)$ signals the breakdown of
our semiclassical analysis using single segment supported $\phi(x)$,
rather than any inconsistencies associated with the stochastic matrix
model.

Therefore, our single support segment solutions (eqs.\ (20)-(22)) are
valid in the $N > \beta$ phase, only for odd multicriticalities $k$.
WKB analysis of the non interacting Fermi gas in the case of pure
gravity ($k = 2$ and a cubic matrix potential)\cite{FD3,AJM} reveals
that in the $N > \beta$ phase, the extension of the semiclassical
particle density of the gas into the complex eigenvalue plane developes
an additional small support segment, with complex conjugate branch
points, in the (double scaled) vicinity of the multicritical end point
$b$ of eq.\ (19).  This second segment has been considered first
in \cite{FD3} as a source of inconsistencies in the definition of
macroscopic loops in the stochastic definition of \cite{MP} for pure
gravity.  However, as has been argued in \cite{AJM}, one should
disregard
this secondary cut (along which the eigenvalue density is complex) by
subtracting it from the definition of macroscopic loop expectations,
since by definition, the fermi particles in the gas have real position
coordinates and their density is real and non-negative.  Indeed,
numerical analysis of the macroscopic loop expectation value (for pure
gravity) reveals no inconsistencies of the type described in \cite{FD3}
at negative cosmological constant\cite{KR}.  Such double segment
particle distributions cannot correspond to collective field
configurations, because they imply  complex field configurations,
which are excluded\footnote{Recall that the guiding principle in
constructing the collective hamiltonian\cite{Sak} is the hermiticity of
the latter, which breaks down for complex collective fields.}.
Moreover,
in such cases the identification of eq.\ (17) with the real part of eq.\
(18) breaks down.

The pecularities of such a collective field distribution, had such a
configuration existed, can be seen by substituting an ansatz for
$\phi(x)$ in the form considered by \cite{FD3} for the case of pure
gravity into eq.\ (17) above.  Doing that, one finds that $\mu_F$ is
necessarily not vanishing and that the resultant scaling of
susceptibility with cosmological constant is that of $k=3$
multicriticality.  Since by simple scaling arguments\cite{FD3,AJM} the
stabilised model retains its $k=2$ scaling behaviour also for negative
cosmological constant, it is clear that such eigenvalue distributions
are
inadmissible.  The fact that we have managed to obtain from $k=2$ to its
irrelevant perturbation $k=3$ (on the sphere) is misleading here, since
eq.\ (17) with non-vanishing $\mu_F$ parameter does not correspond to an
ordinary spherical loop equation (eq.\ (18)).

Some details of the computation described above and some remarks on the
case of $\mu_F \neq 0$ are deferred to the appendix.\\

\noindent{\bf V. \ The Gloomy Fate of KdV Structure at Stabilised Even
Multicriticalities}\\

The source to the powerful KdV structure among scaling correlators in
(unstabilised) matrix models\cite{GrMg,DgSh,BrKz} is the fact that the
Heisenberg algebra may be represented on single fermion orbitals in the
one dimensional gas\cite{BDShS} by coordinate and momentum matrices with
a finite number of off-diagonals\cite{Dg,Br}.  Such a representation may
be realised explicitly only on orbitals which are orthogonal
polynomials relative to a polynomial matrix potential\footnote{These
systems represent the Toda hierarchy before the continuum limit is
taken\cite{MARTI}, which serve as the source for KdV in the continuum.
Had there been another realisation of this Heisenberg algebra that would
have led to the KdV kierarchy only at the double scaling limit by
severely damping faraway off diagonal matrix elements (which we find
hard to believe), it would certainly not have had the  Toda structure.}.
In this section we prove by using very simple arguments, that the exact
ground state of the foreward Fokker Planck hamiltonian in eq.\ (1), is
{\em not} a Slater determinant of orthogonal polynomials relative to a
polynomial matrix potential when $U(\Phi)$ is bottomless.  Thus, as is
clear from the definition of correlators in eq.\ (4) and from the
discussion above the whole KdV structure is lost at even
multicriticalities.

A Slater determinant of orthogonal polynomials relative to a matrix
polynomial potential $V(x)$ is clearly a matrix polar coordinate
representation\cite{BIZ} of the normalisable wave function
\begin{eqnarray}
\Psi_V(\Phi) = \exp [ - \frac{\beta}{2} {\rm Tr} V (\Phi)] \ .      
\end{eqnarray}

Therefore, it is enough to show that $\Psi(\Phi)$ in eq.\ (23) cannot
be an eigenstate of ${\cal H}$ in eq.\ (1) when the original matrix
potential is bottomless.

Assuming the contrary, ${\cal H} \Psi_V = \beta^2 E \Psi_V$, we find
\begin{eqnarray}
\frac{1}{8} {\rm Tr} (U^{\prime^2} - V^{\prime^2}) -
\frac{1}{4\beta} {\rm Tr} \left[(\frac{\partial}{\partial \Phi})
(U^\prime - V^\prime) \right] = E = {\rm const} \ .        
\end{eqnarray}
For $U$ of quartic or a higher degree, the first term on the
l.h.s. of
eq.\ (24) is linear in traces of powers of $\Phi$, while the second term
is bi-linear in these traces. These two terms can never balance one another
to produce a non-vanishing constant E on the r.h.s. of eq.\ (24).
Thus, the latter equation possesses a unique solution, namely,
\begin{eqnarray}
V = U \ , ~~E = 0                    
\end{eqnarray}

This is also the unique solution for cubic potential $U$.
In cases of odd multicriticalities, where $U$ is bounded from below, we
simply recover eq.\ (2) and the discussion following it.  However, at
even multicriticalities, where $U$ is bottomless, eq.\ (25) proves our
claim.

Clearly, the {\em exact} ground state of the Fermi gas in the case of
pure gravity ($k=2$ and cubic $U(\Phi)$) is a Slater determinant of the
first
$N$ eigenstates of a one body hamiltonian with a complicated quartic
potential\cite{MP,KM,AJV}.  These one particle states are obviously not
orthogonal polynomials, since the only Sturm-Liouville operator defined
along the whole real line, whose square integrable eigenstates are orthogonal
polynomials is the one
dimensional harmonic oscillator hamiltonian -- in conformity with our general
proof above.  Whether the {\em exact} ground state of the interacting Fermi
gas at higher even multicriticalities is also a Slater determinant of
one particle states or not is unknown to us, but the fact that these
stabilised models exhibit the correct genus expansion in the double
scaling limit must imply that there exists a rather good Hartree-Fock
approximation to these ground states.

We find it very interesting from the point of view of one dimensional
Fermi systems that the exact ground state of an interacting one
dimensional Fermi gas, corresponding to a stable matrix potential
$U(\Phi)$ in eq.\ (1) is a Slater determinant (which is a trivial result
of ``supersymmetry'' -- i.e., the existence of an equilibrium eigenvalue
distribution under the Langevin evolution).  This behaviour of the
ground state is equivalent ot the existence of the Toda hierarchy
structure mentioned above, but the role of exact excitations of this gas
is unclear.

The possibility of generating a KdV structure at even multicriticalities
in the continuum, despite the absence of orthogonal polynomials in the
discrete case\footnotemark[15] is ruled out by the numerical analysis
contained in \cite{AJM,DALLEY}, at least in the case of pure gravity.

As we have mentioned in the
introduction, another non perturbatively consistent definition of two
dimensional quantum gravity exists, which preserves the KdV structure at
all multicriticalities\cite{DALLEY}.  Which of the two definitions is
the ``correct'' one is still unknown.  It might be possible that the way
to decide on this choice, is to consider both definitions as distinct
phases of the same statistical ensemble of random surfaces, and to
choose the one which produces the lowest (random surface) free energy at
various regions in the scaling parameter space.\\
\pagebreak

\noindent{\bf VI. \ \ A Variational Principle for Orthonormal
Polynomials on the Whole Real Line}\\

In this section we prove that polynomials orthonormal relative to an
arbitrary weight along the whole real line, must obey Hartree-Fock
equations.
The latter reduces to the Schr\"{o}dinger equations for the eigenstates
of an harmonic oscillator (i.e., the case of Hermite polynomials) in the
case of a Gaussian weight, as expected.  To the best of our knowledge,
this result is new in the general theory of orthogonal polynomials.

With no loss of generality we can take the normalisable weight function
as $e^{-\beta U}$.
Consider the expectation value of the hamiltonian ${\cal H}$ of eq.\ (1)
(with $\beta$U being the exponent of the weight function) in the
normalisable state given by eq.\ (23):
\begin{eqnarray}
\beta^2 {\cal E}_0(V) =
\int d^{N^2}  \Phi e^{-\frac{\beta}{2}{\rm Tr}V}
{\cal H}e^{-\frac{\beta}{2}{\rm Tr}V}   /{\cal Z}_V      
\end{eqnarray}
where
\begin{eqnarray}
{\cal Z}_V =
\int d^{N^2}  \Phi e^{-\beta{\rm Tr}V} =
\langle \Psi_V|\Psi_V\rangle < \infty            
\end{eqnarray}
and clearly
\begin{eqnarray}
{\cal E}_0(V) \geq 0      
\end{eqnarray}
for any normalisable $\Psi_V$.

Since we obviously have
\begin{eqnarray}
{\cal E}_0(U) = 0 \ , ~~~ {\cal Z}_U < \infty      \nonumber
\end{eqnarray}
we expect $\Psi_U$ to be an absolute quadratic minimum for
${\cal E}_0(V)$ in function space of the form given by eqs.\ (23) and
(27).  Indeed, for $V$ infinitesimally deviating from $U$
\begin{eqnarray}
V = U + \delta U        
\end{eqnarray}
we find
\begin{eqnarray}
{\cal E}_0(U+\delta U) =
\langle \Psi_U|\frac{1}{8} {\rm Tr} (\delta U^\prime)^2 | \Psi_U
\rangle/
{\cal Z}_U + {\cal O}(\delta U)^3 \geq 0              
\end{eqnarray}
Let $\{v_i(x)\}^\infty_{i=0}$ be the complete set of orthonormal
functions on the real line  of the form
\begin{eqnarray}
v_i(x) = {\cal P}_i(x)\exp(-\frac{\beta}{2}V(x))     
\end{eqnarray}
where $\{{\cal P}_i(x)\}$ are the orthonormal polynomials relative to
$\exp(-\frac{\beta}{2}V(x))$.  Further, let $\{u_i(x)\}^\infty_{i=0}$ be
the $\{v_i\}$ for $V(x) = U(x)$.
Note that $V(x)$ defines the $v_i(x)$ uniquely, and vice-versa  $(V(x)
= - \frac{2}{\beta} \log \left(\frac{v_0(x)}{{\cal P}_0}\right))$.
Therefore \begin{eqnarray}
K_i(x,y) = \frac{\delta U^\prime(x)}{\delta v_i(y)}     
\end{eqnarray}
should be a regular kernel.  Expressing eq.\ (30) in matrix polar
coordinates, using the Slater determinant representation
\begin{eqnarray}
S = \det_{i,j} v_{i-1}(x_j) \ \ ; \ \ 1 \leq i,j \leq N       
\end{eqnarray}
for $\Psi_V$ (and for $\Psi_U$) we find after some trivial manipulations
\begin{eqnarray}
&&{\cal E}(U + \delta U) = \sum^N_{i=1} \int dx v_i^\ast(x)
\left[ - \frac{1}{2\beta^2} \partial^2_x - \frac{1}{4\beta}
U^{\prime\prime}(x)+\frac{1}{8} U^{\prime^2}(x) + \mu_i \right] v_i(x) +
\nonumber \\
&& ~~~~~ - \frac{1}{4\beta} \Sigma_{i,j}^\prime \int dx dy
v^\ast_i(x)v^\ast_j(y)
\frac{U^\prime(x)-U^\prime(y)}{x-y}
[v_i(x)v_j(y)-v_j(x)v_i(y)] \nonumber \\
&& ~~~~~ - \sum^N_{i=1} \mu_i =
\frac{1}{8} \sum^N_{i=1} \int dx
|u_i(x)|^2 (\delta U^\prime(x))^2 + {\cal O}(\delta U)^3  
\end{eqnarray}
Taking the variational derivative of eq.\ (34) with respect to
$v^\ast_i(x)$, at $V = U$, which must vanish as we have discussed above,
we find:
\begin{eqnarray}
&&\delta {\cal E}_0(U + \delta V)/\delta v_i^\ast(x)|_{\delta U = 0}
= \{ [ - \frac{1}{2 \beta^2} \partial^2_x - \frac{1}{4\beta}
U^{\prime\prime} + \frac{1}{8} U^{\prime^2} +          \nonumber \\
&&~~~~~ + \mu_i ] v_i(x) - \frac{1}{2\beta} \Sigma_j^\prime
\int dy v_j^\ast(y)
\frac{U^\prime(x)-U^\prime(y)}{x-y}
[ v_i(x)v_j(y) - v_j(x)v_i(y) ] \}|_{v_i=u_i}
\nonumber    \\
&&~~~~ = \frac{1}{4} \sum^N_{i=1} \int dy |u_i(y)|^2
\delta U^\prime(y) \frac{\delta U^\prime(y)}{\delta
v_i^\ast(x)}|_{\delta U = 0} \equiv 0    \nonumber
\end{eqnarray}
Thus we obtain the desired Hartree-Fock equations.
\begin{eqnarray}
&&\left[ - \frac{1}{2\beta^2} \partial^2_x - \frac{1}{4\beta}
U^{\prime\prime} (x) + \frac{1}{8} U^{\prime^2}(x) + \mu_i \right]
u_i(x) + \nonumber \\
&& - \frac{1}{2\beta} \Sigma_j^\prime \int dy u^*_j(y)
\frac{U^\prime(x) - U^\prime(y)}{x-y}
\left[u_i(x)u_j(y)-u_j(x)u_i(y)\right] \ = 0    
\end{eqnarray}

For an harmonic matrix potential $U = \frac{1}{2}x^2$, eq.\ (35) reduces
after the substitutions $\hbar = 1/(2\beta)$, $y = x/2$ to
\begin{eqnarray}
\left[ - \frac{\hbar^2}{2}\partial^2_y + \frac{1}{2}y^2+\mu_i -
(N+\frac{1}{2})\hbar \right] u_i(y) = 0      
\end{eqnarray}
corresponding to an harmonic oscillator with unit mass and spring
constant.  Therefore identifying the $i$th eigenvalue as $e_i = (N +
\frac{1}{2})\hbar - \mu_i = \hbar(i + \frac{1}{2}$), we find the
appropriate chemical potentials:
\begin{eqnarray}
\mu_i = \hbar(N-i) \ , 
\end{eqnarray}
that is -- $\mu_i$ is the depth of the $i$-th state below the Fermi
level.\\
\pagebreak

\noindent{\bf VII. \ \ Conclusions}\\

We have presented the stochastic stabilisation of matrix models at
arbitrary multicriticality as a possible non perturbative definition of
two dimensional quantum gravity.

The stabilised models exhibit correct double scaling behaviour and
respect the genus expansion to all its orders.  They coincide with ordinary
matrix integrals at stable odd multicriticalities.  They obviously
always lead to real susceptibilities and non perturbatively consistent
continuum results since all correlators of scaling operators are real.
Stochastic stabilisation seems natural also due to the fact that it is
actually equivalent to dynamics of one dimensional interacting Fermi
gas.  Correlation functions in this gas which are stabilised versions of
the original scaling operators are synchronous in Langevin time -- in
perfect agreement with the framework described in \cite{BDShS}.
Collective field formulation of the stochastic matrix models uses by
definition the same jacobian that is used in $c=1$ matrix
models\cite{DJ}, due to the dependence of matrix variables on Langevin
time.  This fact is essential in order to obtain a non-trivial
collective field theory which reproduces the correct scaling behaviour.
We hold this fact also as an argument in favour of stochastic
stabilisation.

We have shown that supersymmetry and its spontaneous breakdown are
irrelevant in these models.  Only the zero fermion sector of the
supersymmetry introduced in \cite{MP} is relevant for our discussion.
This explains the failure of attempts to interpret these models as one
dimensional suerstrings.  We have proven that KdV structure is lost at
even multicriticalities, in contrast to outcomes of another non
perturbative definition of two dimensional quantum gravity\cite{DALLEY}.
Thus stochastic stabilisation defines a consistent summation of the genus
expansion,  and is a natural choice among other possibilities to sum this
series.

Finally, we have proved, as a byproduct of our analysis, that
polynomials orthogonal relative
to an arbitrary weight along the whole real line must obey a system of
Hartree-Fock equations, by using a variational principle associated with
the Fokker-Planck hamiltonian of stochastic stabilisation.\\

\noindent{\bf Acknowledgements}\\

Valuable discussions with A.\ Auerbach, J.\ Avron, D.\ Bar-Moshe, \v{C}.\
Crnkovi\'{c}, A.\ Mann, M.S.\ Marinov, M.\ Moshe, G.\ Parisi, R.\ Plesser,
and A.\ Schwimmer are kindly acknowledged.  I am especially
indebted to M.\ Moshe for his interest and encouragement during this
work.
\pagebreak

\noindent{\bf Appendix}\\

We discuss in this appendix the peculiarities that arise when one looks
for a solution of eq.\ (17) which have additional support segments in
the complex eigenvalue plane (in the case of pure gravity).  We also
comment on the significance of non vanishing $\mu_F$ parameter in that
equation.

We use a cubic potential to describe the $k=2$ point (we follow the
notations of \cite{Neu,JF})
\setcounter{equation}{0}
\setcounter{num}{1}
\def\theequation{\Alph{num}.\arabic{equation}}
\begin{eqnarray}
U_2(x) = - \frac{1}{6} x^3 + \frac{1}{2} x^2 + x
\end{eqnarray}

Following the discussion at the end of section (IV) we take $\phi(x)$ to
be of the form
\begin{eqnarray}
\phi(x) = \frac{C_2}{\pi} \left\{(2+x)(b-x) [(C_r -
x)^2+C_i^2]\right\}^{\frac{1}{2}}
\end{eqnarray}
Here $C_2 = \frac{1}{4}$ as considered in section (IV) while $b, \ C_r$
and $C_i$ are real parameters.  The branch points $b, \ C_r \pm iC_i$ of
$\phi(x)$ are expected to lie in the vicinity of $x = 2$ in the complex
eigenvalue plane.

Substituting eqs.\ (A.1) and (A.2) into eq.\ (17) we obtain
\begin{eqnarray}
\frac{1}{2}\pi^2\phi(x)^2 + \frac{1}{8} U^\prime_2(x)^2 - \frac{1}{2}
\frac{N}{\beta} (1 - \frac{x+P}{2}) - \mu_F = 0           
\end{eqnarray}
where we have used eq.\ (15) and defined the puncture expectation value
$P$ as  $\frac{\beta}{N} \int^b_{-2} y \phi(y)dy.$

The l.h.s. of eq.\ (A.3) is a cubic polynomial and the vanishing
conditions of its coefficients give rise to four equations in the six
unknowns $b, \ C_r, \ C_i, \ \frac{N}{\beta}, \ P$ and $\mu_F$ which are
supplemented by eq.\ (15) and the definitions of $P$.  Clearly, we must
have $\mu_F \neq 0$, otherwise $\phi(x)$ cannot have a second pair of
branch points, as we know from our explicit construction in section
(IV).

Solving eq.\ (A.3) we immediately obtain the following relations among
the unknown parameters:
\begin{eqnarray}
&&b-2 = 2(2-C_r) \nonumber \\
&&C_i^2 = 3 (2-C_r)^2 \nonumber \\
&&1 - \frac{N}{\beta} = (2-C_r)^3
\end{eqnarray}
and
\begin{eqnarray}
2(2-Cr)^3 + \frac{1}{4} - 2\mu_F-\frac{1}{2}[(2-C_r)^3 - 1]P=0  \ . 
\end{eqnarray}

Clearly, eqs.\ (A.4) imply that the susceptibility $b-2$ scales in the
cosmological constant $1-\frac{N}{\beta}$ as
\begin{eqnarray}
b-2 = 2(1-\frac{N}{\beta})^{\frac{1}{3}}
\end{eqnarray}
typical of the $k=3$ multicritical point.

The result in eq.\ (A.6) does not imply that we have managed to flow from
the $k=2$ point to its irrelevant deformation $k=3$, since eq.\ (17)
with $\mu_F\neq0$ does not correspond to the ordinary planar loop
equation for $k=2$ \ \cite{Neu}.

It has been shown in \cite{JVT,RVT} that a non vanishing $\mu_F$
implies
spontaneous supersymmetry breaking already on the sphere -- which we
actually interpret only as the fact that we see runaway solutions of the
Langevin equation already on the sphere.

We  also note here that if one tries to interpret eq.\ (17) with $\mu_F
\neq 0 $ as equivalent to a planar loop equation in the zero dimensional
matrix model, the latter must contain an anomalous surface term in the
Dyson-Schwinger equation\cite{Neu} related to constant shifts of the
matrix variables $\Phi \rightarrow \Phi+\epsilon$:
\begin{eqnarray}
\int d^{N^2} \Phi \frac{\partial}{\partial \Phi_{ab}}
\left[e^{-\beta{\rm Tr}U} \left(\frac{1}{z-\Phi}\right)_{ab}\right] =
2\mu_F\int d^{N^2}\Phi e^{-\beta{\rm Tr}U} \cdot
\frac{1}{\beta} {\rm Tr} \delta(z-\Phi)  
\end{eqnarray}
This equation seems to hold to all orders in the $1/N$ expansion.

The appearance of the surface term proportional to $\mu_F$ in eq.\
(A.7), and therefore associated with runaway solutions of the Langevin
equation, already at the spherical level might be related to similar
observations made in \cite{MPL}.
\pagebreak


\begin{thebibliography}{99}
\bibitem{GrMg} D.\ Gross and A.\ Migdal, Phys.\ Rev.\ Lett.\ {\bf 64}
(1990) 717; Nucl.\ Phys.\ {\bf B 340} (1990) 333.
\bibitem{DgSh} M.\ Douglas and S.\ Shenker, Nucl.\ Phys.\ {\bf B 335}
(1990) 635.
\bibitem{BDShS} T.\ Banks, M.\ Douglas, N.\ Seiberg and S.\ Shenker,
Phys.\ Lett.\ {\bf B 238} (1990) 279.
\bibitem{BrKz} E.\ Br\'{e}zin and V.\ Kazakov, Phys.\ Lett.\ {\bf B 236}
(1990) 144.
\bibitem{Br} E.\ Br\'{e}zin
in Proc. of the 8th Jerusalem Winter School
for Theoretical Physics on ``Two Dimensional Quantum Gravity and Random
Surfaces'' December 1990-January 1991.  D.J.\ Gross, T.\ Piran and S.\
Weinberg (editors), World Scientific 1992.
\bibitem{Mg} A.A.\  Migdal
in Proc. of the 8th Jerusalem Winter School
for Theoretical Physics on ``Two Dimensional Quantum Gravity and Random
Surfaces'' December 1990-January 1991.  D.J.\ Gross, T.\ Piran and S.\
Weinberg (editors), World Scientific 1992.
\bibitem{DgSeiSh} M.\ Douglas, N.\ Seiberg and S.\ Shenker, Phys.\
Lett.\ {\bf B 244} (1990) 381.
\bibitem{Sh} S.H.\ Shenker, Rutgers preprint RU-90-47 (August 1990),
presented at the Carg\'{e}se Workshop on Random surfaces, quantum
gravity and strings (Carg\'{e}se, France, May 1990).
\bibitem{GinZJ} P.\ Ginsparg and J.\ Zinn-Justin, Saclay preprint
SPhT/90-140, Los Alamos preprint LA-UR-90-3687 (October 1990),
presented at the Carg\'{e}se Workshop on Random surfaces, quantum
gravity and strings (Carg\'{e}se, France, May 1990).
\bibitem{FD1} F.\ David, Mod.\ Phys.\ Lett.\ {\bf A 5} (1990) 1019.
\bibitem{FD2} F.\ David, Nucl.\ Phys.\ {\bf B 348} (1991) 507.
\bibitem{FD3} F.\ David, Saclay preprint SPhT/90-178 (May 1990),
presented at the Carg\'{e}se Workshop on Random surfaces, quantum
gravity and strings (Carg\'{e}se, France, May 1990).
\bibitem{FD4} F.\ David
in Proc. of the 8th Jerusalem Winter School
for Theoretical Physics on ``Two Dimensional Quantum Gravity and Random
Surfaces'' December 1990-January 1991.  D.J.\ Gross, T.\ Piran and S.\
Weinberg (editors), World Scientific 1992.
\bibitem{BrDgKzSh} E.\ Br\'{e}zin, M.R.\ Douglas, V.A.\ Kazakov and
S.H.\ Shenker, Phys.\ Lett.\ {\bf B 237} (1990) 43;\\
C. Crnkovi\'{c}, P.\ Ginsparg and G.\ Moore, Phys.\ Lett.\ {\bf B 237}
(1990) 196.
\bibitem{ChLk} S.\ Chaudhuri and J.D.\ Lykken, Nucl.\ Phys.\ {\bf B 367}
(1991) 614.         
\bibitem{MP} E.\ Marinari and G.\ Parisi, Phys.\ Lett.\ {\bf B 240}
(1990) 375.
\bibitem{GrHal} J.\ Greensite and M.B.\ Halpern, Nucl.\ Phys.\ {\bf B
242} (1984) 167.
\bibitem{KM} M.\ Karliner and A.\ Migdal, Mod.\ Phys.\ Lett.\ {\bf A 5}
(1990) 2565.
\bibitem{AJV} J.\ Ambjo\hspace{-0.20cm}/rn, J.\ Greensite and S.\
Varsted, Phys.\ Lett. {\bf B 249} (1990) 411.\\
J.\ Ambjo\hspace{-0.20cm}/rn and J.\ Greensite, Phys.\ Lett.\ {\bf B
254} (190) 66.
\bibitem{MSGV} J.L.\ Miramontes, J.S.\ Guillen and M.A.H.\ Vozmediano,
Phys.\ Lett.\ {\bf B 253} (1991) 38.\\
J. Gonz\'{a}lez and M.A.H.\ Vozmediano, Phys.\ Lett.\ {\bf B 258} (1991)
55.
\bibitem{AD} A.\ Dabholkar, Nucl.\ Phys.\ {\bf B 368} (1992) 293.
\bibitem{MSG} J.L.\ Miramontes and J.\ S\'{a}nchez Guill\'{e}n, in
``Proceedings of the Workshop on Random Surfaces and 2D Quantum
Gravity'' Barcelona 1991, Nucl.\ Phys.\ {\bf B (Proc.\ Suppl.) 66}
(1991) 1.\\
J.L.\ Miramontes and J.\ S\'{a}nchez Guill\'{e}n, CERN preprint
CERN-TH-6323/91.
\bibitem{Neu} H.\ Neuberger, Nuc.\ Phys.\ {\bf B 352} (1991) 689.
\bibitem{BIZ} D.\ Bessis, C.\ Itzykson and J.-B.\ Zuber, Adv.\ Appl.\
Math.\ {\bf 1} (1980) 109.    
\bibitem{BPIZ} E.\ Br\'{e}zin, C.\ Itzykson, G.\ Parisi and J.-B.\
Zuber, Commun.\ Math.\ Phys.\ {\bf 59} (1978) 35.
\bibitem{Kaz} V.A.\ Kazakov, Mod.\ Phys.\ Lett.\ {\bf A.4} (1989) 2125.
\bibitem{JF} J.\ Feinberg, Phys.\ Lett.\ {\bf B 281} (1991) 225.
\bibitem{Sak} B.\ Sakita, Quantum theory of many-variable systems and
fields (World Scientific, Singapore, 1985);\\
A.\ Jevicki and B.\ Sakita, Nucl.\ Phys.\ {\bf B 165} (1980) 511; {\bf B
185} (1981) 89.    
\bibitem{DJ} S.R.\ Das and A.\ Jevicki, Mod.\ Phys.\ Lett.\ {\bf A 5}
(1990) 1639.
\bibitem{JR} A.\ Jevicki and J.P.\ Rodrigues, Phys.\ Lett.\ {\bf B 268}
(1991) 53. 
\bibitem{JVT} A.J. van Tonder, Ph.D. Thesis, Witwatersrand Univ., March
1992, CLNS-92-01.
\bibitem{RVT} J.P.\ Rodrigues and A.J.\ van Tonder, Witwatersrand
Preprint CLNS-92-02.
\bibitem{SBM} S.\ Ben-Menahem SLAC Preprint, SLAC-PUB-5377, February
1992 (A preliminary version has been circulated in November 1990). 
\bibitem{JCDA} J.D.\ Cohn and S.P.\ de Alwis, Nucl.\ Phys.\ {\bf B 368}
(1992) 79.
\bibitem{Jev} A.\ Jevicki, Brown preprint BROWN-HET-777 (October 1990).
\bibitem{Dg} M.R.\ Douglas, Phys.\ Lett.\ {\bf B 238} (1990) 176.
\bibitem{Loop} R.\ Dijkgraaf, H.\ Verlinde and E.\ Verlinde, Nucl.\
Phys.\ {\bf 348} (1991) 435;\\
M.\ Fukuma, H.\ Kawai and R.\ Nakayama, Intern.\ J.\ Mod.\ Phys.\ {\bf A
6} (1991) 1385.   
\bibitem{DALLEY} S.\ Dalley, C.\ Johnson and T.R.\ Morris, Nucl.\ Phys.\
{\bf B 368} (1992) 625-654; ibid 655-670; Phys.\ Lett.\ {\bf B 291}
(1992) 11 (and references therein).
\bibitem{MORRIS} T.R.\ Morris, Nuc.l.\ Phys.\ {\bf B 356} (1991) 703.
\bibitem{AJM} J.\ Ambjo\hspace{-0.20cm}/rn, C.V.\ Johnson and T.R.\
Morris, Nucl.\ Phys.\ {\bf B 374} (1992) 496.
\bibitem{PaBk} G. Parisi, Statistical Field Theory (Addison-Wesley,
Reading, MA, 1988) ch.\ 19. 
\bibitem{ZJBk} J.\ Zinn-Justin, Quantum Field Theory and Critical
Phenomena (Clarendon, Oxford, 1989) ch.\ 3.
\bibitem{HCT} S.R.\ Das, A.\ Dhar, A.M.\ Sengupta and S.R.\ Wadia, Mod.\
Phys.\ Lett. {\bf A 5} (1990) 1041;\\
D.J.\ Gross and M.J.\ Newman, Phys.\ Lett.\ {\bf B 266} (1991) 291;\\
Yu.\ Makeenko and G.W.\ Semenoff, Mod.\ Phys.\ Lett.\ {\bf A 6} (1991)
3455.   
\bibitem{GrKl} D.J.\ Gross and I.R.\ Klebanov, Nucl.\ Phys.\ {\bf B
352} (1991) 671.
\bibitem{MPL} E.\ Marinari and G.\ Parisi, Phys.\ Lett.\  {\bf B 247}
(1990) 537.
\bibitem{Witten} E.\ Witten, Nucl.\ Phys.\ {\bf B 185} (1981) 513.
\bibitem{DIVA} J.\ Distler and C.\ Vafa, Princeton preprint -- PUPT-1212
(May 1990), Presented at the Carg\'{e}se Workshop on Random Surfaces,
Quantum Gravity and Strings. Mod.\ Phys.\ Lett.\ {\bf A 6} (1991) 259.
\bibitem{CDL} S.\ Chaudhuri, H.\ Dykstra and J.\ Lykken, Fermilab
preprint Fermi-conf-91/190-T (July 1991). Mod.\ Phys.\ Lett.\ {\bf A6}
(1991) 1665.
\bibitem{CIT} C.-I.\ Tan, Phys.\ Rev.\ {\bf D 45} (1992) 2862.
\bibitem{DMP} R.\ Dijkgraaf, G.\ Moore and R.\ Plesser, Yale Preprint
YCTP-P22-92 (hepth@xxx/9208031).  
\bibitem{JF1} J.\ Feinberg, work in progress.
\bibitem{KR} M.\ Karliner, A.\ Migdal and B.\ Rusakov -- private communication.
\bibitem{MARTI} E.J.\ Martinec, Commun.\ Math.\ Phys.\ {\bf 138} (1991)
437; and Rutgers Preprint RU-91-51 presented at the 1991 Trieste Spring
School.




\end{thebibliography}
\end{document}